\documentclass[entropy,article,submit,pdftex,moreauthors]{Definitions/mdpi}
\preto{\abstractkeywords}{\nolinenumbers}
\firstpage{1}
\pubvolume{0}
\issuenum{0}
\articlenumber{0}
\pubyear{0}
\copyrightyear{0}
\datereceived{0}
\daterevised{0}
\dateaccepted{0}
\datepublished{0}
\hreflink{https://doi.org/}
\newcommand{\RNumb}{\mathbb{R}}
\newcommand{\CNumb}{\mathbb{C}}

\newcommand{\UnitOp}{\hat{1\kern-4.75pt 1}} % operator unit with hat
\newcommand{\MatUnit}{1\kern-4.75pt 1} % matrix unit

\newcommand{\Norm}[1]{\|\kern.3ex#1\kern.3ex \|} % norm macro

\newcommand{\Bra}[1]{\langle #1 \vert}
\newcommand{\Ket}[1]{\vert #1 \rangle}
\newcommand{\BraKet}[2]{\langle #1 \vert #2 \rangle}
\newcommand{\Prob}{\mathrm{Prob}\,}

\newcommand{\EOp}{\mathsf{E}\kern-1pt\llap{$\vert$}}
\newcommand{\WOp}{\mathsf{W}\kern-1pt\llap{$-$}}
\newcommand{\FOp}{\mathsf{F}\kern-1pt\llap{$\vert$}}
\newcommand{\StateSpace}[1]{\mathcal{#1}} % state space

\Title{
Delayed Choice Phenomena in the Projection Evolution Model}
\TitleCitation{
Delayed Choice Phenomena in the Projection Evolution Model}

\Author{
Marek Góźdź $^{1}$\orcidA{},
Andrzej Góźdź $^{2}$\orcidB{} and
Krzysztof Lider $^{2}$}
\AuthorNames{Marek Góźdź, Andrzej Góźdź, Krzysztof Lider}
\AuthorCitation{Góźdź, M.; Góźdź, A.; Lider, K.}
\address{
$^{1}$ \quad Institute of Computer Science and Mathematics,
Maria Curie-Skłodowska University,
ul. Akademicka 9, 20-033 Lublin, Poland;
mgozdz@kft.umcs.lublin.pl
\\
$^{2}$ \quad Institute of Physics,
Maria Curie-Skłodowska University,
pl. Marii Curie-Skłodowskiej 1, 20-031 Lublin, Poland;
andrzej.gozdz@umcs.lublin.pl
}
\corres{Correspondence: mgozdz@kft.umcs.lublin.pl}

\abstract{In the Schr\"odinger evolution of a~quantum state time enters as
  a~real parameter representing the coordinate. In a~more consistent approach
  time should be defined as a~quantum observable, with the evolution taking
  place in a~four-dimensional spacetime. This is possible in the projection
  evolution model in which the wave function is defined in both space and
  time. This allows to construct the time operator and discuss the temporal
  structure of quantum processes. In this paper we discuss a~photon travelling
  through a~Mach-Zehnder interferometer, focusing the description on the
  temporal profile of the wave function. We show that in this approach the
  delayed-choice experiments can be explained by the temporal overlap of the
  photon and the devices in the interferometer.}

\keyword{foundations of quantum mechanics; quantum time; projection evolution
  model}

\PACS{03.65.-w, 03.65.Ca, 03.65.Ta}

\begin{document}
\maketitle
%%%%%%%%%%%%%%%%%%%%%%%%%%%%%%%%%%%%%%%%%%%%%%%%%%%%%%%%%%%%%%%%%%%%%%%%

%%%%%%%%%%%%%%%%%%%%%%%%%%%%%%%%%%%%%%%%%%%%%%%%%%%%%%%%%%%%%%%%%%%%%%%%
\section{Introduction}
%%%%%%%%%%%%%%%%%%%%%%%%%%%%%%%%%%%%%%%%%%%%%%%%%%%%%%%%%%%%%%%%%%%%%%%%

In the standard approach to non-relativistic quantum mechanics the time
evolution of a quantum state $\psi$ is given by the equation
\begin{equation}
  \psi(t,x) = U(t) \psi(x,0),
  \label{eq:SchEv}
\end{equation}
where $U(t)$ is a unitary evolution operator generated by a Hamiltonian $H$
and $t$ denotes time. In the case of time independent Hamiltonian
$U(t)= \exp(-i H t)$, where here and in the following we set $\hbar=1$.
Despite being useful in many special cases, time in this formulation is a
classical parameter rather than a~quantum observable. It follows that the wave
function $\psi$ is ``quantized'' in space but not in time, which poses
interpretational problems with the space-time formulation of the quantum
theory. On the other hand the Pauli theorem forbids to represent time as
a~self-adjoint operator in the space $L^2(\RNumb^3,d^3 x)$, so that
Eq.~(\ref{eq:SchEv}) cannot be corrected by simply replacing $t$ with its
operator form. 

In 1978 J.A.~Wheeler proposed a~thought experiment \cite{Wheeler} in which
a~double-slit setup was changed --- giving the possibility to register either
the wave-like or the particle-like nature of the photon --- right before the
detection. He argued that the detected nature of the photon should be in full
agreement with the setup, thus that the quantum particle cannot be described
by the classical concept of motion. This idea has been experimentally proven
first by Walborn \textit{et.al.} \cite{exp-Walborn}, then by Aspect
\textit{et.al.} \cite{exp-Aspect} and others. The explanation involved the
wave function of the particle being spatially wide, reaching to the location
where the changes in the setup were introduced. On the other hand the temporal
double-slit experiments \cite{exp-Houser,exp-Lindner,exp-Tirole} strongly
suggested the need of treating time as a~quantum observable. Having time as
a~quantum observable would provide a~natural explanation for the
delayed-choice experiments.

Many attempts have been made to reformulate quantum mechanics and properly
describe time in the model, including the works by Ahoronov and Bohm
\cite{th-Aharonov}, Rosenbaum \cite{th-Rosenbaum}, Grot, Rovelli, and Tate
\cite{th-Grot}, Olkhovsky \cite{th-Olkhovsky}, Kijowski \cite{th-Kijowski} and
many other. In these papers the authors managed to recreate the Schr\"odinger
evolution (\ref{eq:SchEv}) usually by postulating different interpretations of
the parameter $t$ and its connection to the actual time. Even though Galapon
\cite{th-Galapon} showed that for a~discrete spectrum Hamiltonian
a~self-adjoint time operator can be constructed, its consistent definition was
still missing.

In this work we will use the projection evolution approach \cite{pev-opis} to
discuss the delayed-choice experiment in the context of the temporal
interaction. In this model time $t=x^0$ is the component of the four-position
$x=(x^0,\vec x)$ and is properly represented in the form of a self-adjoint
operator with eigenvalues being the time coordinates of the system. The time
evolution of a~quantum state is then given by a series of mappings from the
current state onto the space of possible subsequent states. This generic model
allows to describe unitary and non-unitary evolution scenarios as well as
cases in which the Hamiltonian-based desciption is problematic. We focus our
discussion on the delayed-choice scenario and show, that the temporal part of
the wave function can be used to describe photon behaviour in such setup in
a~natural way.

In Sec.~\ref{sec:PEv} we describe the basic idea behind the projection
evolution. We do not elaborate on all the details here, so for an extensive
background and more technical discussion see Ref.~\cite{pev-opis} and
references therein. In Sec.~\ref{sec:MZ} we apply the projection evolution to
a photon travelling through a Mach-Zehnder interferometer, in which the
beamsplitters are inserted and removed at certain instances of time. We
calculate in Sec.~\ref{sec:examples} the detection probabilities for both
detectors, taking into account the temporal width of the photon and its
temporal overlap with the beamsplitters.

%%%%%%%%%%%%%%%%%%%%%%%%%%%%%%%%%%%%%%%%%%%%%%%%%%%%%%%%%%%%%%%%%%%%%%%%
\section{The projection evolution}
\label{sec:PEv}
%%%%%%%%%%%%%%%%%%%%%%%%%%%%%%%%%%%%%%%%%%%%%%%%%%%%%%%%%%%%%%%%%%%%%%%%

We describe the time evolution of a~quantum state using the projection evolution
(PEv) formalism \cite{pev-opis}. We start from a~four-dimensional formulation in
which time is a~quantum observable similar to the three spatial coordinates. It
follows that the wave function $\psi=\psi(x)$ depends on all four coordinates
$x=(x^0,x^1,x^2,x^3)$ and the spacetime position operator $\hat{x}^\mu$ acts as
a multiplication operator, i.e., $\hat{x}^\mu \psi(x)= x^\mu \psi(x)$.  Such
approach allows to describe in a~consistent way different equations of motion,
like the Schr\"odinger equation, Dirac equation, Klein-Gordon equation, and
other.

The four-dimensional wave function gives the probability of finding the object
at some given space-time location, i.e., it has certain width in the spatial and
in the temporal direction. This allows to define the uncertainties
$\Delta\vec{x}$ and $\Delta{t}$, localize the object in space and time
intervals, and discuss the time structure of quantum processes.

The PEv method is rooted in the earlier works by Choi \cite{pev-Choi} and
Krauss \cite{pev-Krauss} about quantum operations. Since for a~consistent
model time has to be a~quantum observable, it cannot serve as a~parameter of
the evolution any longer. We introduce therefore an index $\tau$ which labels
subsequent steps of the quantum evolution in spacetime observing, that at each
evolution step the quantum state is dependent on all four components of the
position vector. The most general evolution of a~quantum state is given by an
operator
\begin{equation}
  \EOp: \StateSpace{K}_\tau \to \StateSpace{K}_{\tau+1}
\end{equation}
which maps the initial Hilbert space $\StateSpace{K}_\tau$ at the evolution
step $\tau$ onto the Hilbert space $\StateSpace{K}_{\tau+1}$ at the next step
of the evolution. For the process described in this paper, the Hilbert space
remains the same during the whole evolution, i.e.,
$\StateSpace{K} := \StateSpace{K}_\tau = \StateSpace{K}_{\tau+1}$. 

The evolution operators $\EOp$ can have different forms, leading to different
evolution equations. If $\EOp$ are unitary operators we end up with the
Schr\"odinger-like evolution. If $\EOp$ are projection operators, the space
$\StateSpace{K}$ consists of all possible states on which the projections are
made according to the appropriate probability distribution. In every case the
quantum state at the step $\tau+1$ is given by the normalized action of the
evolution operators $\EOp_{\tau+1}$ at this step onto the previous quantum
state,
\begin{equation}
  \psi(\tau+1;t,x) = \frac{\EOp_{\tau+1} \psi(\tau;t,x)}{\Norm{\EOp_{\tau+1}
  \psi(\tau;t,x)}}.
\end{equation}

An evolution process of a~quantum state has to be described by modifications
of the state $\psi$ at different steps $\tau$. A~temporal localization of the
system may happen during the evolution, in which case the state is projected
onto the time axis and the time coordinate takes a~definite value. In the case
of no time projection, the time coordinate is smeared over certain interval,
depending on the width of the time component of the wave function. It is
therefore important to remember that $\tau$ is not time and that a~definite
temporal coordinate can be obtained by projecting the state onto the time
axis.

In our approach the wave function describes the position of the particle in
time and in space alike, i.e., at spacetime. The spatial uncertainty
$\Delta\vec{x}$ comes in pair with the momentum uncertainty
$\Delta\vec{p}$. The temporal uncertainty $\Delta x^0 = \Delta t$ pairs with
the temporal momentum uncertainty $\Delta p_0$, where
$p_0 = i\frac{\partial}{\partial t}$. Here the $p_0$ operator gets the meaning
of the momentum along the temporal axis and its sign represents the arrow of
time, i.e., the direction of the evolution of the state. The localization of
the particle in space and time comes as the result of interactions acting
similarly to measurements; a~quantum clock can be built based upon this idea
\cite{pev-zegar}.

%%%%%%%%%%%%%%%%%%%%%%%%%%%%%%%%%%%%%%%%%%%%%%%%%%%%%%%%%%%%%%%%%%%%%%%%
\section{The Mach-Zehnder interferometer}
\label{sec:MZ}
%%%%%%%%%%%%%%%%%%%%%%%%%%%%%%%%%%%%%%%%%%%%%%%%%%%%%%%%%%%%%%%%%%%%%%%%

\begin{figure}[th]
  \centering
  \includegraphics[scale=0.75]{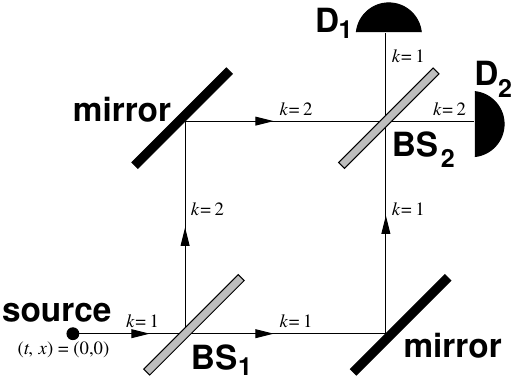}
  \caption{The Mach-Zehnder interferometer}
  \label{fig:MZinter}
\end{figure}

The Mach-Zehnder interferometer consists of two beamsplitters, two mirrors,
and two detectors, as shown in Fig.~\ref{fig:MZinter}.

%%%%%%%%%%%%%%%%%%%%%%%%%%%%%%%%%%%%%%%%%%%%%%%%%%%%%%%%%%%%%%%%%%%%%%
%%%%%%%%%%%%%%%%%%%%%%%%%%%%%%%%%%%%%%%%%%%%%%%%%%%%%%%%%%%%%%%%%%%%%%

The state space of the interferometer consists of two quantum channels based
on functions dependent on spacetime coordinates. Assuming channel separation
the simplified description contains two orthogonal channels in
a~two-dimensional spacetime, as indicated in Fig.~\ref{fig:MZinter}. In this
case we can reduce the full state space to a~simpler one
$\StateSpace{K}=L^2(\RNumb^2,dt\,dx) \otimes \CNumb^2$. It implies that every
state $\bar{\Psi} \in \StateSpace{K}$ can be written as
\begin{equation}
\label{StateVec2}
\bar{\Psi}(t,x):= \BraKet{t,x}{\bar{\Psi}}
= \psi(t,x) \otimes \sum_{k=1}^2 c_k \Ket{k},
\end{equation}
where $c_k\in\CNumb$, $|c_1|^2+|c_2|^2=1$, and the scalar product of two
states $\bar{\Psi}'$ and $\bar{\Psi}$ is given by
\begin{equation}
  \label{ScProdK2}
  \BraKet{\bar{\Psi}'}{\bar{\Psi}}=
  \int_{\RNumb^2} dt dx \ \psi'(t,x)^\star \psi(t,x)
  \sum_{k=1}^2 {c'_k}^\star c_k.
\end{equation}
This approach does not change the qualitative behaviour of our system.

The photon wave function can be constructed by introducing electromagnetic
vector fields $\mathbf{E}$ and $\mathbf{H}$ in agreement with the Maxwell
equations \cite{Szpikowski}. An alternative way is to solve the Proca
equation, as the resulting scalar and vector fields obey in the massless case
the Maxwell conditions. Following these methods one gets (up to normalization)
a~unitary temporal part $e^{i\omega_t t}$ and the spatial part in the form of
a~spherical Bessel function
$\frac{1}{r}\sqrt{\frac{\omega r}{c}} J_{j+1/2}(\frac{\omega}{c}r)$.

Following the full solutions for the quantized electromagnetic field
\cite{Szpikowski}, we use for simplification the ``one dimensional photon''
(1D-photon) wave function $\BraKet{t,x}{\tau;\gamma}=\gamma(\tau;t,x)$ at the
evolution step $\tau$ in a~separable form:
$\gamma(\tau;t,x)=\gamma^t(\tau;t)\gamma^x(\tau;x)$, where the temporal part
of the photon is given by the Fourier transform of an appropriate frequency
profile $a(\omega_t)$
\begin{equation}
  \label{foton-czas}
  \Ket{\tau;\gamma^t} \to \gamma^t(\tau;t)
  = \frac{1}{\sqrt{2\pi}} \int_{\RNumb} d\omega_t a(\omega_t)
  e^{i\omega_t (t-\alpha(\tau)) },
\end{equation}
with $\omega_t$ describing the temporal width of the photon's profile. This
form leads to a~uniform probability distribution in time. In the lowest order
of the Bessel function, the spatial part can be written as
\begin{equation}
  \label{gammaXState}
  \Ket{\tau;\gamma^x} \to \gamma^x(\tau;x)
  = \sqrt{\frac{\omega_x}{\pi}}
  \frac{ \sin(\omega_x (x-\alpha(\tau)))}{\omega_x(x-\alpha(\tau))},
\end{equation}
where the parameter $\omega_x$ represents the spatial width of the photon's
profile.

To avoid unnecessary complication of the description, we consider
a~semiclassical motion of the photon as a shift in spacetime. Following this
approach, the maximum of the state \eqref{gammaXState} shifts during the
evolution along the $x$ axis, simulating the one-dimensional semiclassical
motion of the photon. These shifts are given by $c \alpha(\tau)$, which for
setting the speed of light $c=1$ simplifies to $\alpha(\tau)$. This also
represents the spatial and temporal localizations of the maxima of the photon
wave function at the evolution step $\tau$.

In the following, we write only these evolution operators which describe our
process and we omit all other alternatives. The evolution of the state of the
photon can be described by a~sequence of interactions with the elements of the
interferometer and propagation between them. We may list the following seven
steps: at $\tau=\tau_0$ we start with the input state $\Ket{\bar{\Psi}}_{in}$;
at $\tau=\tau_1$ the photon reaches the first entrance channel of the
interferometer; at $\tau=\tau_2$ the photon may interact with the first
beamsplitter BS$_1$; at $\tau=\tau_3$ the photon reaches the mirrors and at
$\tau=\tau_4$ interacts with them; at $\tau=\tau_5$ it is shifted to the
second beamsplitter BS$_2$, and at $\tau=\tau_6$ it may interact with it; at
$\tau=\tau_7$ the photon arrives to the detectors and is measured.

The photon is produced in the source at $\tau_0$ in the first channel, so the
first evolution operator reads:
\begin{equation}
  \label{EOp0}
  \EOp(\tau_0;1)=
  \Ket{\tau_0;\gamma}\Bra{\tau_0;\gamma} \otimes \Ket{1}\Bra{1}.
\end{equation}
Acting on the input state $\Ket{\bar{\Psi}}_{in}$ this operator determines the
initial state of the photon as
\begin{equation}
  \label{PsiTau0}
  \Ket{\bar{\Psi}(\tau_0)}=\Ket{\tau_0;\gamma} \otimes \Ket{1}
\end{equation}
Here we assume the initial shift $\alpha(\tau_0)=0$.

As it was mentioned above the free motion of the photon is described by
a~simple shift in the spacetime along the world line $x=t$. This implies the
following evolution operator:
\begin{equation}
  \label{EOpF}
  \EOp(\tau;F)= \hat{T}(\alpha(\tau))  \otimes \UnitOp,
\end{equation}
where $\hat{T}(\alpha)$ is the spacetime translation operator in
$\StateSpace{K}$. This means that
\begin{equation}
  \label{StateTransl}
  \hat{T}(\alpha) \bar{\Psi}(t,x)= \bar{\Psi}(t-\alpha,x-\alpha).
\end{equation}
Reaching the beamsplitter BS$_1$ the photon state is
\begin{equation}
  \label{EOp1}
  \Ket{\bar{\Psi}(\tau_1)}
  =\EOp(\tau_1;F) \Ket{\tau_0;\gamma} \otimes \Ket{1} 
  = \Ket{\tau_1;\gamma} \otimes\Ket{1}.
\end{equation}
In general the beamsplitter could modify the spacetime profile of the photon.
As this plays no role for our analysis we will not take it into account and
represent the beamsplitters by the evolution operators
\begin{equation}
  \label{EOpBS}
  \EOp(\tau;\mathrm{BS}_\ell):=
  P(\Omega_{\ell})\otimes \hat{B}_\ell
  + (1-P(\Omega_{\ell}))\otimes \UnitOp,
\end{equation}
where
\begin{equation}
  \label{EOpProj}
  P(\Omega):= 
  \int_{\RNumb^2}dtdx\, \Ket{t,x} \delta((t,x) \in \Omega) \Bra{t,x}
\end{equation}
projects onto the subset $\Omega$ of the spacetime positions. The set
$\Omega_{\ell}$, $\ell=1,2$, describes the spacetime profile of the
$\ell^{\rm th}$ beamspliter, i.e., it determines when and where the
beamspliter is present in the system. We have introduced the Boolean function
$\delta(\mathcal{C})$ defined as follows: $\delta(\mathcal{C})=1$ if the
condition $\mathcal{C}$ is fulfilled, otherwise this function is equal to
zero.

The operators $\hat{B}_\ell$ mix channels and are represented for BS$_1$ and
BS$_2$ in the channel basis $\{\Ket{1},\Ket{2}\}$ by:
\begin{eqnarray}
  && \hat{B}_1 \Ket{1} = \frac{1}{\sqrt{2}} ( \Ket{1} + \Ket{2} ), \quad
     \hat{B}_1 \Ket{2} = \frac{1}{\sqrt{2}} (-\Ket{1} + \Ket{2} ), \\
  && \hat{B}_2 \Ket{1} = \frac{1}{\sqrt{2}} ( \Ket{1} - \Ket{2} ), \quad
     \hat{B}_2 \Ket{2} = \frac{1}{\sqrt{2}} ( \Ket{1} + \Ket{2} ).
\end{eqnarray}
After the first beamsplitter the normalized state is given by
\begin{equation}
  \label{BS1State}
  \bar{\Psi}(\tau_2)(t,x)
  =\delta((t,x) \in \Omega_{1}) \BraKet{t,x}{\tau_1;\gamma}
  \otimes \frac{1}{\sqrt{2}} \left(\Ket{1}+\Ket{2} \right)
  + \delta((t,x) \not\in \Omega_{1}) 
  \BraKet{t,x}{\tau_1;\gamma} \otimes \Ket{1}.
\end{equation}
Again, the free evolution shifts the resulting state to the next part of the
interferometer and the state entering the mirrors is given by
\begin{eqnarray}
  \label{AfterBS1S}
  && \bar{\Psi}(\tau_3)(t,x)
     = \hat{T}(\alpha(\tau_3)-\alpha(\tau_1)) \bar{\Psi}(\tau_2)(t,x) 
     \nonumber\\
  && = \delta((t-(\alpha_3-\alpha_1),x-(\alpha_3-\alpha_1)) \in \Omega_{1})
     \BraKet{t,x}{\tau_3;\gamma}
     \otimes \frac{1}{\sqrt{2}} \left(\Ket{1}+\Ket{2} \right)
     \nonumber\\
  && + \delta((t-(\alpha_3-\alpha_1),x-(\alpha_3-\alpha_1)) \not\in \Omega_{1})
     \BraKet{t,x}{\tau_3;\gamma} \otimes \Ket{1}, 
\end{eqnarray}
where $\alpha_k \equiv \alpha(\tau_k)$.

The mirrors are large enough to prevent the photons from passing around
them. Let us assume that the mirrors can modify phases of the photon in each
arm of the interferomer, independently. The evolution operator $\EOp(\tau;M)$
is diagonal,
\begin{equation}
  \label{EOpM}
  \EOp(\tau;M(\kappa_1,\kappa_2)) = \UnitOp \otimes \hat{M}(\kappa_1,\kappa_2),
\end{equation}
where $\hat{M}(\kappa_1,\kappa_2)$ acts on the basic channel states as
\begin{equation}
  \label{Mirrors}
  \hat{M}(\kappa_1,\kappa_2)\Ket{k}= e^{i\kappa_k}\Ket{k}.
\end{equation}
The mirrors change the previous state as follows:
\begin{eqnarray}
  \label{MState}
  && \bar{\Psi}(\tau_4;t,x)
     = \delta((t-(\alpha_3-\alpha_1),x-(\alpha_3-\alpha_1))
     \in \Omega_{1}) \BraKet{t,x}{\tau_3;\gamma}
     \otimes \frac{1}{\sqrt{2}}
     \left(e^{i\kappa_1}\Ket{1} + e^{i\kappa_2}\Ket{2} \right) 
     \nonumber\\
  && + \delta((t-(\alpha_3-\alpha_1),x-(\alpha_3-\alpha_1))
     \not\in \Omega_{1}) \BraKet{t,x}{\tau_3;\gamma} 
     \otimes e^{i\kappa_1}\Ket{1}.
\end{eqnarray}
The free motion to the beamsplitter BS$_2$ prepares its input state as 
\begin{eqnarray}
  \label{BS2State}
  && \bar{\Psi}(\tau_5;t,x)
     = \hat{T}(\alpha(\tau_5)-\alpha(\tau_3)) \bar{\Psi}(\tau_4;t,x) 
     \nonumber\\
  && = \delta((t-(\alpha_5-\alpha_1),x-(\alpha_5-\alpha_1)) \in \Omega_{1})
     \BraKet{t,x}{\tau_5;\gamma} \otimes \frac{1}{\sqrt{2}}
     \left(e^{i\kappa_1}\Ket{1}+e^{i\kappa_2}\Ket{2} \right) 
     \nonumber\\ 
  && + \delta((t-(\alpha_5-\alpha_1),x-(\alpha_5-\alpha_1)) \not\in \Omega_{1})
     \BraKet{t,x}{\tau_5;\gamma} \otimes e^{i\kappa_1}\Ket{1} \phantom{xx}.
\end{eqnarray}
At the evolution step $\tau_6$ the beamsplitter BS$_2$ mixes channels which
results in a~possible interference:
\begin{eqnarray}
  \label{EOpBS2}
  && \bar{\Psi}(\tau_6;t,x) 
     = \delta((t-(\alpha_5-\alpha_1),x-(\alpha_5-\alpha_1)) \in \Omega_{1})
     \delta((t,x) \in \Omega_{2}) \BraKet{t,x}{\tau_5;\gamma} 
     \nonumber\\
  && \otimes \frac{1}{2}\{ (e^{i\kappa_1} + e^{i\kappa_2})\Ket{1}
     + (-e^{i\kappa_1} + e^{i\kappa_2})\Ket{2} \} 
     \nonumber\\
  && + \delta((t-(\alpha_5-\alpha_1),x-(\alpha_5-\alpha_1)) \not\in \Omega_{1})
     \delta((t,x) \in \Omega_{2}) \BraKet{t,x}{\tau_5;\gamma}
     \otimes \frac{e^{i\kappa_1}}{\sqrt{2}} (\Ket{1}-\Ket{2})
     \nonumber\\
  && +\delta((t-(\alpha_5-\alpha_1),x-(\alpha_5-\alpha_1)) \in \Omega_{1})
     \delta((t,x) \not\in \Omega_{2}) \BraKet{t,x}{\tau_5;\gamma}
     \otimes \frac{1}{\sqrt{2}} (e^{i\kappa_1}\Ket{1}+e^{i\kappa_2}\Ket{2})
     \nonumber\\
  && + \delta((t-(\alpha_5-\alpha_1),x-(\alpha_5-\alpha_1)) \not\in \Omega_{1})
     \delta((t,x) \not\in \Omega_{2})
     \BraKet{t,x}{\tau_5;\gamma} \otimes e^{i\kappa_1} \Ket{1}.
\end{eqnarray}
The state \eqref{EOpBS2} describes four independent basic scenarios. For all
of them the next evolution step is a free motion of the photon to the
detectors. It is described by the vector
$\Ket{\bar{\Psi}(\tau_7)}= \hat{T}(\alpha(\tau_7)-\alpha(\tau_5))
\Ket{\bar{\Psi}(\tau_6)}$ which has the same structure as \eqref{EOpBS2}. The
vector $\Ket{\bar{\Psi}(\tau_7)}$ is the final state of the photon before its
measurement by the detectors. The most important information is the
probability distribution of detecting the photon in the detector D$_k$ which
monitors the channel $\Ket{k}$.

In our case each detector is represented the projection operator
$P(\mathrm{D}_k)$ which localizes the photon in the time interval
$(\bar{t}-\epsilon_t/2,\bar{t}+\epsilon_t/2)$ and at the spatial position in
the interval $(\bar{x}-\epsilon_x/2,\bar{x}+\epsilon_x/2)$. Similarly to the
mirrors we assume that the photon cannot pass the detectors undetected. The
evolution operators read
\begin{equation}
  \label{Detect}
  P(\mathrm{D}_k) =
  \int_{\bar{t}-\epsilon_t/2}^{\bar{t}+\epsilon_t/2} dt\,
  \int_{\bar{x}-\epsilon_x/2}^{\bar{x}+\epsilon_x/2} dx\,
  \Ket{t,x} \otimes \Ket{k}  \Bra{k} \otimes\Bra{t,x},
\end{equation}
where $\Ket{t,x}$ are eigenstates of the time and position operators in
$L^2(\RNumb^2,dt\,dx)$ represented by the standard multiplication type
operators: $\hat{t}f(t,x)=t f(t,x)$ and $\hat{x}f(t,x)=x f(t,x)$
\cite{pev-opis}.

The required probability of finding the photon in the detector $k$ at time
$t \in(\bar{t}-\epsilon_t/2,\bar{t}+\epsilon_t/2)$ is given by the standard
formula
\begin{equation}
  \label{ProbD}
  \Prob(D_k; \bar{t},\bar{x})
  = \Bra{\bar{\Psi}(\tau_7)} P(\mathrm{D}_k) \Ket{\bar{\Psi}(\tau_7)}
  = \int_{\bar{t}-\epsilon_t/2}^{\bar{t}+\epsilon_t/2} dt\,
  \int_{\bar{x}-\epsilon_x/2}^{\bar{x}+\epsilon_x/2} dx\,
  |(\Bra{t,x}\otimes \Bra{k})\Ket{\bar{\Psi}(\tau_7)}|^2.
\end{equation}
Using appropriately shifted states \eqref{EOpBS2} the spacetime representation
of the measured state is 
\begin{eqnarray}
  \label{Psi7}
  && (\Bra{t,x}\otimes \Bra{k})\Ket{\bar{\Psi}(\tau_7)} 
     \nonumber\\
  && = \delta((t-(\alpha_7-\alpha_1),x-(\alpha_7-\alpha_1)) \in \Omega_{1})
     \delta((t,x) \in \Omega_{2}) \gamma(\tau_7;t,x) 
     \nonumber\\
  && \frac{1}{2}\{ (e^{i\kappa_1} + e^{i\kappa_2}) \delta_{k1}
     + (-e^{i\kappa_1} + e^{i\kappa_2})\delta_{k2} \} 
     \nonumber\\
  && + \delta((t-(\alpha_7-\alpha_1),x-(\alpha_7-\alpha_1)) \not\in \Omega_{1})
     \delta((t,x) \in \Omega_{2}) \gamma(\tau_7;t,x)
     \frac{e^{i\kappa_1}}{\sqrt{2}} (\delta_{k1}-\delta_{k2})
     \nonumber\\
  && +\delta((t-(\alpha_7-\alpha_1),x-(\alpha_7-\alpha_1)) \in \Omega_{1})
     \delta((t,x) \not\in \Omega_{2}) \gamma(\tau_7;t,x)
     \frac{1}{\sqrt{2}} (e^{i\kappa_1}\delta_{k1}+e^{i\kappa_2}\delta_{k2})
     \nonumber\\ 
  && + \delta((t-(\alpha_7-\alpha_1),x-(\alpha_7-\alpha_1)) \not\in \Omega_{1})
     \delta((t,x) \not\in \Omega_{2})
     \gamma(\tau_7;t,x) e^{i\kappa_1} \delta_{k1}.
\end{eqnarray}
Again, we get four independent alternatives instead of sixteen combinations of
matrix elements. This is due to the fact that the Boolean funcion
$\delta(\mathcal{C})$ is a projector, i.e.,
$\delta(\text{true})^2 = \delta(\text{true})=1$ and
$\delta(\text{true})\delta(\text{false})=0$. The density probability function
in \eqref{ProbD} can be now written as
\begin{eqnarray}
  \label{PDensPsi7}
  && |(\Bra{t,x}\otimes \Bra{k})\Ket{\bar{\Psi}(\tau_7)}|^2 =
     |\gamma(\tau_7;t,x)|^2   
     \nonumber\\
  && \Big\{ 
     \delta((t-(\alpha_7-\alpha_1),x-(\alpha_7-\alpha_1)) \in \Omega_{1})
     \delta((t-(\alpha_7-\alpha_5),x-(\alpha_7-\alpha_5)) \in \Omega_{2}) 
     \nonumber\\
  && \times \frac{1}{2} [(1+\cos(\kappa_1-\kappa_2)) \delta_{k1}
     + (1-\cos(\kappa_1-\kappa_2))\delta_{k2}] 
     \nonumber\\
  && + \frac{1}{2} 
     \delta((t-(\alpha_7-\alpha_1),x-(\alpha_7-\alpha_1)) \not\in \Omega_{1})
     \delta((t-(\alpha_7-\alpha_5),x-(\alpha_7-\alpha_5)) \in \Omega_{2}) 
     (\delta_{k1} + \delta_{k2})
     \nonumber\\
  && + \frac{1}{2}  
     \delta((t-(\alpha_7-\alpha_1),x-(\alpha_7-\alpha_1)) \in \Omega_{1}) 
     \delta((t-(\alpha_7-\alpha_5),x-(\alpha_7-\alpha_5)) \not\in \Omega_{2})
     (\delta_{k1} + \delta_{k2})
     \nonumber\\
  && + 
     \delta((t-(\alpha_7-\alpha_1),x-(\alpha_7-\alpha_1)) \not\in \Omega_{1})
     \delta((t-(\alpha_7-\alpha_5),x-(\alpha_7-\alpha_5)) \not\in \Omega_{2}) 
     \delta_{k1}  
     \Big\}.
\end{eqnarray}
One needs to rememeber that $\alpha_k=\alpha(\tau_k)$ represents the distance
from the source of the photon to the appropriate point in the interferometer.

%%%%%%%%%%%%%%%%%%%%%%%%%%%%%%%%%%%%%%%%%%%%%%%%%%%%%%%%%%%%%%%%%
\section{Discussion}
\label{sec:examples}
%%%%%%%%%%%%%%%%%%%%%%%%%%%%%%%%%%%%%%%%%%%%%%%%%%%%%%%%%%%%%%%%%

We assume that the photon enters the interferometer at $(t,x)=(0,0)$,
and that the first beamsplitter, the mirrors, the second beamsplitter,
and the detectors are all 5 units of space apart, ie., BS$_1$ can be
reached at $x=5$, the mirrors at $x=10$, BS$_2$ at $x=15$, and the
detectors at $x=20$. In the following we assume $\kappa_1=\kappa_2=\pi$.

If there is only one beamsplitter present in the system, the wave
function of the photon is statistically split into both channels,
reaching both of the detectors with probability $\frac{1}{2}$. If there
is no beamsplitter present, only the detector D$_1$ will detect the
photon. If both beamsplitters interact with the photon, the interference
will enhance the signal in D$_1$ and destroy the signal in D$_2$.

If the beamsplitters are removed and inserted in the setup, the photon
may or may not react to the change. This is called the delayed-choice
experiment and it is usually explained invoking the spatial width of the
wave function. If the photon has a~non-zero overlap with the
beamsplitter, it will modify its state in accordance with the changes
made to the setup. In our approach, however, time enters the photon's
position four-vector and should also be considered. Below we discuss
some delayed-choice scenarios focusing on the temporal interaction and
show, that the results are fully compatible with the expectations.

\subsection{The symmetric case}

The behaviour of the photon will depend on its spatial and temporal
profiles. Let us start with the photon wave function in the shape of
a~box, with sharp boundaries, both in space and time. The normalized
step function centered around $x_0$ is given by:
\begin{equation}
  \mathrm{step}(x-x_0) = \left\{ 
    \begin{array}{ll}
      \frac{1}{\Delta_x}, & 
      x_0 - \frac{\Delta_x}{2} < x < x_0 + \frac{\Delta_x}{2} \\
      0, & \mathrm{otherwise}
    \end{array} \right.
\end{equation}
and similarly for time
\begin{equation}
  \mathrm{step}(t-t_0) = \left\{ 
    \begin{array}{ll}
      \frac{1}{\Delta_t}, & 
      t_0 - \frac{\Delta_t}{2} < t < t_0 + \frac{\Delta_t}{2} \\
      0, & \mathrm{otherwise}
    \end{array} \right..
\end{equation}
The photon will interact with everything from within the boxes, so if
the beamsplitter appears earlier or later than the position $(t_0,x_0)$
but has a~non-zero overlap with the photon's profile, it will affect the
particle. The rectangular box, even though artificial, serves as a~good
illustration of the mechanism of the photon interaction with the
experimental setup. Due to the existence of sharp boundaries the
spacetime region occupied by the particle is always clearly defined,

In a~more physically viable case the temporal rectangular box can be
replaced by a~normalized Gaussian
\begin{equation}
  \gamma^t(t) = \frac{1}{\sqrt{2\pi}\omega_t} e^{-\frac{t^2}{2\omega_t^2}}
\label{gammaT3}
\end{equation}
and the spatial part by the lowest order of Eq.~(\ref{gammaXState}).
This photon is symmetric both in the forward and in the backward
direction of time. It means that the changes made to the system before
and after photon has reached the required spacetime point can still
alter the quantum state. This remark is valid for all the elements in
the interferometer: the beamsplitters, the mirrors, and even the
detectors; the projection evolution model does not distinguish between
them. 

\begin{figure}[th]
  \centering
  \includegraphics[width=0.65\textwidth]{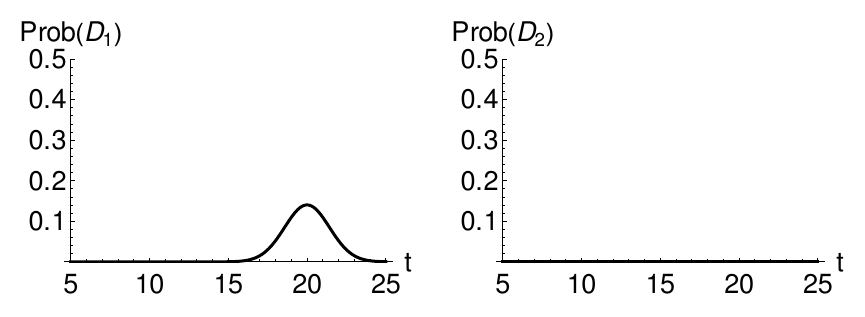} \\
  \includegraphics[width=0.65\textwidth]{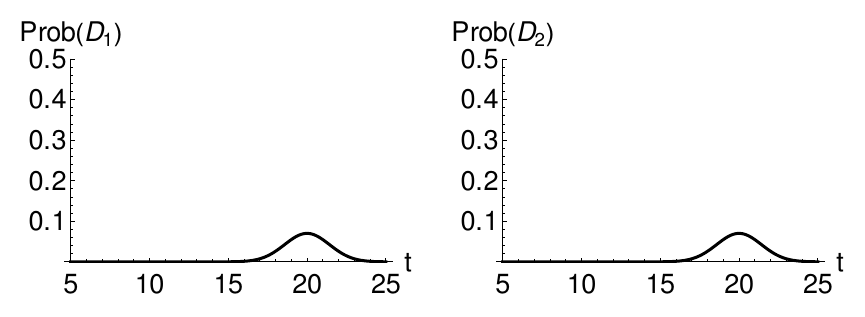}
  \caption{The symmetric photon time profile. The detection
    probabilities for none or both beamsplitters present in the setup
    (upper figure); for only one of the beamsplitters present in the
    setup (lower figure).}
  \label{fig:wyniki-sym}
\end{figure}

In Fig.~\ref{fig:wyniki-sym} the detection probabilities for detectors
D$_1$ and D$_2$ are presented. One notices that the maximum detection
probability appears for $(t_0,x_0)=(20,20)$, i.e., for the spacetime
location of the detector. The same result will be obtained for the
photon being treated as a~classical point-like particle. The position of
the classical particle is represented by the position of the maximal
detection probability of the photon, which in the case of symmetric
profile coincides with the position of the global maximum.

%%%%%%%%%%%%%%%%%%%%%%%%%%%%%%%%%%%%%%%%%%%%%%%%%%%%%%%%%%%%%%%%%%%%%%
%%%%%%%%%%%%%%%%%%%%%%%%%%%%%%%%%%%%%%%%%%%%%%%%%%%%%%%%%%%%%%%%%%%%%%

\subsection{The asymmetric case}

The other possibility is the asymmetric case in which the photon has
a~strong temporal maximum with a~tail directed forwards or backwards in
time. It may be given in the form
\begin{equation}
  \gamma^t(t) = (\omega_t^2 (\pm t)) e^{-\omega_t (\pm t)}
\label{gammaT}
\end{equation}
and represent the situation in which the particle either senses the time
interval before its maximum or can probe later times.

\begin{figure}[th]
  \centering
  \includegraphics[width=0.65\textwidth]{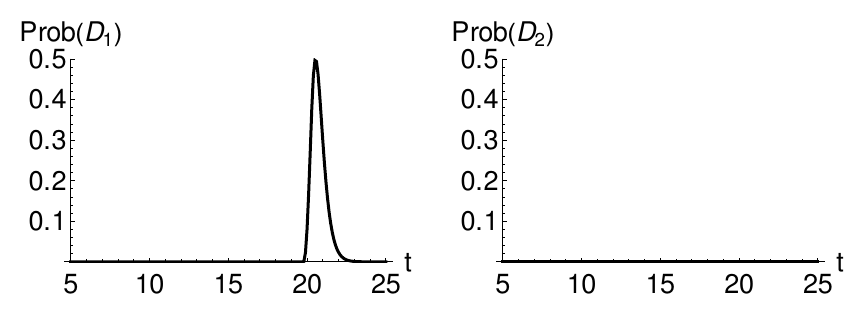} \\
  \includegraphics[width=0.65\textwidth]{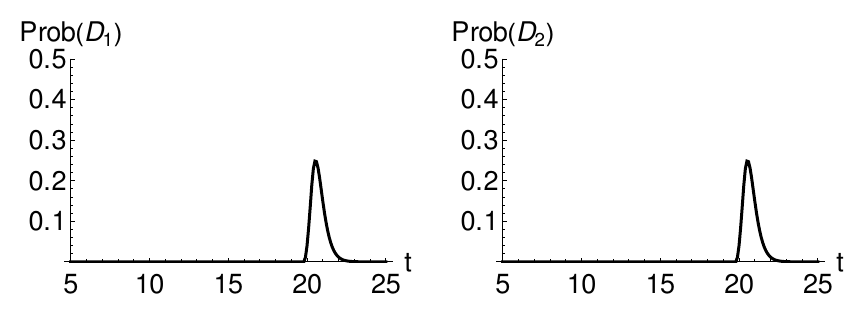}
  \caption{The asymmetric photon time profile. The detection
    probabilities for none or both beamsplitters present in the setup
    (upper figure); for only one of the beamsplitters present in the
    setup (lower figure).}
  \label{fig:wyniki-przod}
\end{figure}

\begin{figure}[th]
  \centering
  \includegraphics[width=0.65\textwidth]{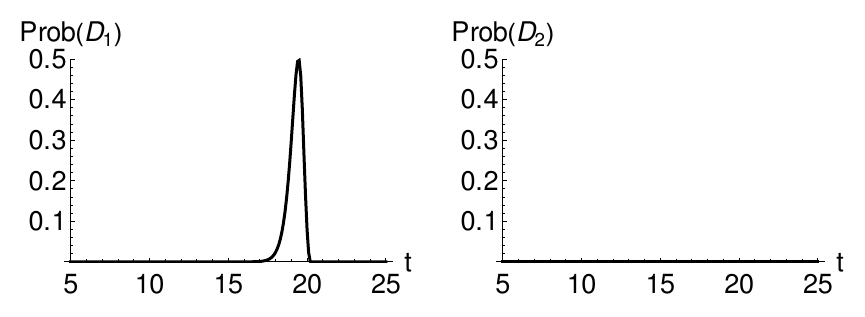} \\
  \includegraphics[width=0.65\textwidth]{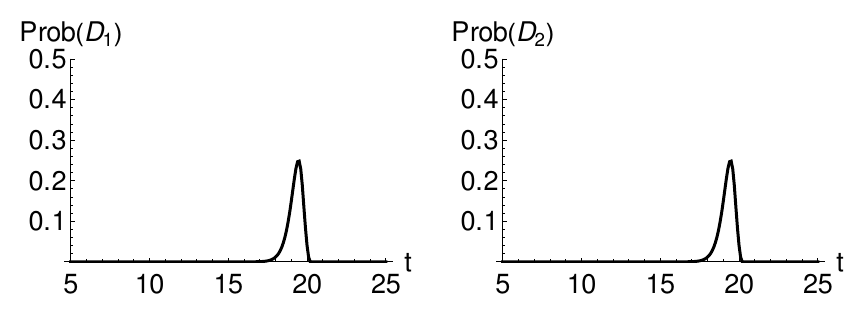}
  \caption{The asymmetric photon time profile. The detection
    probabilities for none or both beamsplitters present in the setup
    (upper figure); for only one of the beamsplitters present in the
    setup (lower figure).}
  \label{fig:wyniki-tyl}
\end{figure}

In these cases the changes made to the interferometer will only affect
the photon, if they appear in the correct time interval. If the photon
has a~backwards time tail, it will react to all the changes made before
it has reached the given spacetime point. It means that it will know
about all the manipulations of the beamsplitters before reaching their
spacetime localization. Contrary, the photon with a~time tail directed
forward in time will react to the changes made after he has passed the
given spacetime location. This is valid not only for the experimental
setup, but also for the detectors, which means that a~photon with
a~forward time tail has a~probability to be detected earlier than the
photon with a~backward time tail. This is ilustrated in
Figs.~\ref{fig:wyniki-przod} and \ref{fig:wyniki-tyl}.

%%%%%%%%%%%%%%%%%%%%%%%%%%%%%%%%%%%%%%%%%%%%%%%%%%%%%%%%%%%%%%%%%%%%%%
%%%%%%%%%%%%%%%%%%%%%%%%%%%%%%%%%%%%%%%%%%%%%%%%%%%%%%%%%%%%%%%%%%%%%%

%%%%%%%%%%%%%%%%%%%%%%%%%%%%%%%%%%%%%%%%%%%%%%%%%%%%%%%%%%%%%%%%%%%%%%
%%%%%%%%%%%%%%%%%%%%%%%%%%%%%%%%%%%%%%%%%%%%%%%%%%%%%%%%%%%%%%%%%%%%%%

\subsection{Other examples}

We will present below few special cases in which the beamsplitters are
inserted in the interferometer before or after photon has reached their
spatial location. In all of the cases the temporal parts of the photon
wave function give the possibility of interaction with the
beamsplitter. This, alongside with the spatial interaction, gives
a~valid explanation of the delayed-choice experiments.

\begin{figure}[th]
  \centering
  \includegraphics[width=0.65\textwidth]{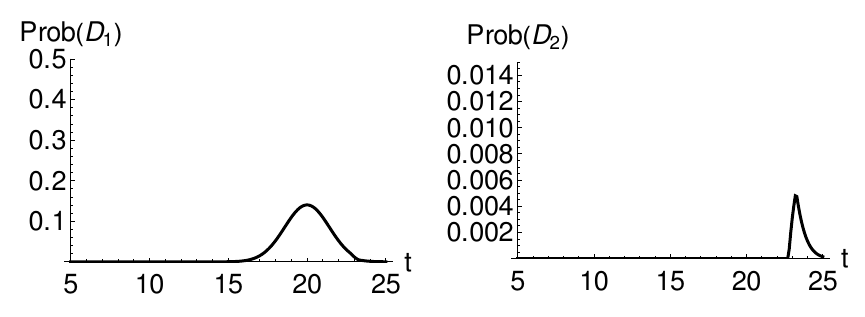}
  \caption{The detection probabilities in Scenario 1.}
  \label{fig:KL2Gauss}
\end{figure}

\textit{Scenario 1:} In the first scenario BS$_1$ is absent and BS$_2$
is present for $18 \le t \le 21$. The temporal part of the photon is
a~Gaussian. In this case a~small detection probability appears in the
second detector for times greater than $t=20$ (see
Fig.~\ref{fig:KL2Gauss})

%%%%%%%%%%%%%%%%%%%%%%%%%%%%%%%%%%%%%%%%%%%%%%%%%%%%%%%%%%%%%%%%%%%%%%
%%%%%%%%%%%%%%%%%%%%%%%%%%%%%%%%%%%%%%%%%%%%%%%%%%%%%%%%%%%%%%%%%%%%%%

\begin{figure}[th]
  \centering
  \includegraphics[width=0.65\textwidth]{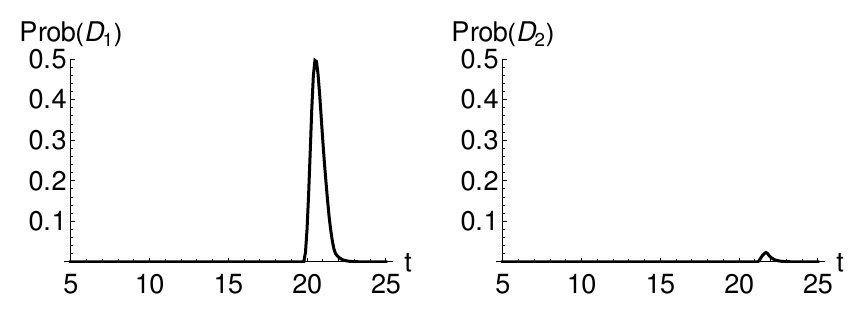} \\
  \includegraphics[width=0.65\textwidth]{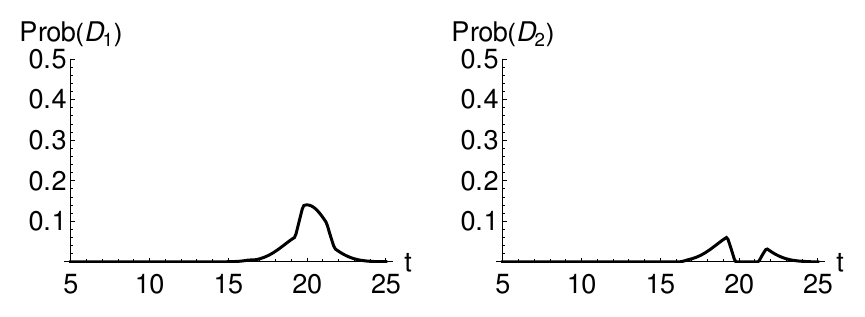}
  \caption{The detection probabilities in Scenario 2 for a~forward
    directed temporal profile (upper figure) and a~Gaussian (lower figure).}
  \label{fig:KL1PPrzod}
\end{figure}

\textit{Scenario 2:} The first beamsplitter is present before the photon
can reach its spatial position, $1.5 \le t \le 4.5$. The second
beamsplitter appears later, $16.5 \le t \le 19.5$. The photon has
a~temporal tail directed forwards. In this case the second detector gets
a~small probability of detection (see Fig.~\ref{fig:KL1PPrzod}). We
notice that if the photon had the temporal tail in the backwards
direction, $D_2$ would not react as the photon could not react with BS$_2$.

%%%%%%%%%%%%%%%%%%%%%%%%%%%%%%%%%%%%%%%%%%%%%%%%%%%%%%%%%%%%%%%%%%%%%%
%%%%%%%%%%%%%%%%%%%%%%%%%%%%%%%%%%%%%%%%%%%%%%%%%%%%%%%%%%%%%%%%%%%%%%

\begin{figure}[th]
  \centering
  \includegraphics[width=0.65\textwidth]{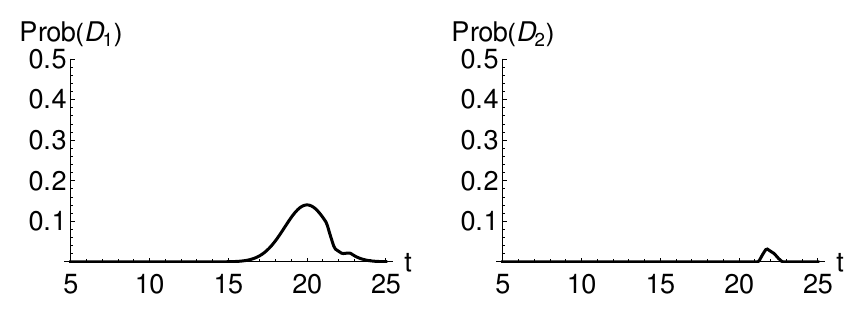}
  \caption{The detection probabilities in Scenario 3.}
  \label{fig:MG1Gauss}
\end{figure}

\textit{Scenario 3:} The BS$_1$ is present at $6.5 \le t \le 7.5$, the
BS$_2$ at $6.5 \le t \le 9.5$, so we have a~situation when only one,
then both, and again only one beamsplitter is present in the
system. A~Gaussian photon will have the detection probabilities as shown
in Fig.~\ref{fig:MG1Gauss}

%%%%%%%%%%%%%%%%%%%%%%%%%%%%%%%%%%%%%%%%%%%%%%%%%%%%%%%%%%%%%%%%%%%%%%
%%%%%%%%%%%%%%%%%%%%%%%%%%%%%%%%%%%%%%%%%%%%%%%%%%%%%%%%%%%%%%%%%%%%%%

%%%%%%%%%%%%%%%%%%%%%%%%%%%%%%%%%%%%%%%%%%%%%%%%%%%%%%%%%%%%%%%%%%%%%%%%
\section{Closing remarks}
%%%%%%%%%%%%%%%%%%%%%%%%%%%%%%%%%%%%%%%%%%%%%%%%%%%%%%%%%%%%%%%%%%%%%%%%

The temporal part of the wave function provides a~natural way to describe the
time structure of quantum events like the temporal interference,
time-of-arrival, nuclear fission and fusion processes, time structure of
elementary particles interactions and many more. In this paper we have shown
that the delayed-choice experiment on the example of a~Mach-Zehnder
interferometer can be successfully described by the temporal interaction. This
result fully supports the expectations of Wheeler, and does not need the
introduction of retrocausality. This misconception comes from the fact, that
causality is often treated in a~classical way, as if time for quantum systems
was the same parametric time as for the macroscopic objects. Treating time as
a~quantum observable not only makes the model consistent, but also alters the
notion of causality for quantum systems -- one may speak about causality only
between points in which the object was localized on the time axis. Between
them the object is smeared over some time interval, exactly like in the case
of spatial localization. In the presented here example of the interferometer
the photon was not localized until detected, having the possibility to react
to the changes in the setup.

The free parameter $\omega_t$ controls the width of the temporal profile. Up
to our best knowledge there is no direct experimental data on this subject,
but in normal conditions we expect the temporal widths to be very small. This
has, however, to be verified.

%%%%%%%%%%%%%%%%%%%%%%%%%%%%%%%%%%%%%%%%%%%%%%%%%%%%%%%%%%%%%%%%%%%%%%%%
\bibliography{art43-interferometr.bib}

@Book{Wheeler,
author = "{J.A.~Wheeler}",
editor = "{A.R.~Marlow}",
title = "{Mathematical Foundations of Quantum Theory}",
publisher = "Academic Press",
year = 1978
}

@Article{exp-Walborn,
author = "{S.P.~Walborn, M.O.~Terra Cunha, S.~P\'adua, C.H.~Monken}",
title = "{A double-slit quantum eraser}",
journal = "{Phys. Rev. A}",
year = 2002,
volume = 65,
pages = 033818
}

@Article{exp-Aspect,
author = "{V.~Jacques, E.~Wu, F.~Grosshans, F.~Treussart,
           P.~Grangier, A.~Aspect, J-F.~Roch}",
title = "{Experimental Realization of Wheeler's
          Delayed-Choice Gedanken Experiment}",
journal = "Science",
year = 2007,
volume = 315,
pages = "966--968"
}

@Article{exp-Houser,
author = "{U.~Houser, W.~Neuwirth, N.~Thesen}",
title = "{Time-dependent modulation of the probability amplitude
          of single photons}",
journal = "{Phys. Lett. A}",
year = 1974,
volume = 49,
pages = "57--58"
}

@Article{exp-Lindner,
author = "{F.~Lindner, M.~Sch\"atzel, H.~Walther, A.~Baltu{\v s}ka,
           E.~Goulielmakis, F.~Krausz, D.~Milo{\v s}evi\'c, D.~Bauer,
           W.~Becker, G.~Paulus}",
title = "{Attosecond Double-Slit Experiment}",
journal = "{Phys. Rev. Lett.}",
year = 2005,
volume = 95,
pages = 040401
}

@Article{exp-Tirole,
author = "{R.~Tirole, S.~Vezzoli, E.~Galiffi, I.~Robertson, D.~Maurice,
           B.~Tilmann, S.A.~Maier, J.B.~Pendry, R.~Sapienza}",
title = "{Double-slit time diffraction at optical frequencies}",
journal = "{Nature Physics}",
year = 2023,
volume = 19,
pages = "999--1002"
}

@Article{th-Aharonov,
author = "{Y.~Aharonov, D.~Bohm}",
title = "{Time in the Quantum Theory and the Uncertainty Relation
          for Time and Energy}",
journal = "{Phys. Rev.}",
year = 1961,
volume = 122,
pages = 1649
}

@Article{th-Rosenbaum,
author = "{D.M.~Rosenbaum}",
title = "{Super Hilbert Space and the Quantum‐Mechanical Time Operators}",
journal = "{J. Math. Phys.}",
year = 1969,
volume = 10,
pages = "1127--1144"
}

@Article{th-Grot,
author = "{N.N.~Grot, C.~Rovelli, R.S.~Tate}",
title = "{Time of arrival in quantum mechanics}",
journal = "{Phys. Rev. A}",
year = 1996,
volume = 54,
pages = 4676
}

@Article{th-Olkhovsky,
author = "{V.S.~Olkhovsky, E.~Recami, A.J.~Gerasimchuk}",
title = "{Time operator in quantum mechanics}",
journal = "{Il Nuovo Cimento A}",
year = 1974,
volume = 22,
pages = "263--278"
}

@Article{th-Kijowski,
author = "{J.~Kijowski}",
title = "{On the time operator in quantum mechanics and the heisenberg
          uncertainty relation for energy and time}",
journal = "{Rep. Math. Phys.}",
year = 1974,
volume = 6,
pages = "361--386"
}

@Article{th-Galapon,
author = "{E.~Galapon}",
title = "{Pauli's theorem and quantum canonical pairs: the consistency
          of a bounded, self–adjoint time operator canonically conjugate
          to a Hamiltonian with non–empty point spectrum}",
journal = "{Proc. R. Soc. Lond. A}",
year = 2002,
volume = 458,
pages = "451--472"
}

@Article{pev-opis,
author = "{A.~Góźdź, M.~Góźdź, A.~Pędrak}",
title = "{Quantum Time and Quantum Evolution}",
journal = "{Universe}",
year = 2023,
volume = 9,
pages = 256
}

@Article{pev-zegar,
author = "{A.~Góźdź, M.~Góźdź}",
title = "{Quantum Clock in the Projection Evolution Formalism}",
journal = "{Universe}",
year = 2024,
volume = 10,
pages = 116
}

@Article{pev-Choi,
author = "{M.~Choi}",
title = "{Completely positive linear maps on complex matrices}",
journal = "{Lin. Alg. App.}",
year = 1975,
volume = 10,
pages = "285--290"
}

@Book{pev-Krauss,
author = "{K.~Krauss}",
title = "{States, Effects and Operations:
          Fundamental Notions of Quantum Theory}",
publisher = {Springer: Berlin/Heidelberg, Germany},
year = 1983
}

@Book{Szpikowski,
author = "{S.~Szpikowski}",
title = "{Podstawy mechaniki kwantowej}",
publisher = "{Wydawnictwo UMCS, Poland}",
note = "{(in Polish)}",
year = 2011
}
\end{document}